\documentclass[oribibl,orivec]{llncs}
\usepackage[T1]{fontenc}
\usepackage{amsmath,amssymb}
\usepackage{microtype}
\usepackage[round,authoryear]{natbib}
\usepackage[hidelinks]{hyperref}

\title{\texorpdfstring{Two Primes Suffice for\\Definable Graph Canonisation}{Two Primes Suffice for Definable Graph Canonisation}}
\titlerunning{Two Primes Suffice for Definable Graph Canonisation}
\author{Vaani Goenka\\{\normalfont\small Ashoka University}\\[.65em]
Aalok Thakkar\\{\normalfont\small Vachani School of Advanced Computing, Ashoka University}}
\institute{}
\hypersetup{pdftitle={Two Primes Suffice for Definable Graph Canonisation},
  pdfauthor={Vaani Goenka and Aalok Thakkar}}
\hypersetup{bookmarksdepth=2}

\spnewtheorem*{namedtheorem}{Theorem}{\bfseries}{\itshape}
\spnewtheorem*{namedlemma}{Lemma}{\bfseries}{\itshape}
\spnewtheorem*{namedcorollary}{Corollary}{\bfseries}{\itshape}
\spnewtheorem*{namedproposition}{Proposition}{\bfseries}{\itshape}

\newcommand{\Aut}{\operatorname{Aut}}
\newcommand{\CF}{\operatorname{cf}}
\newcommand{\ModP}{\mathrm{Mod}_p\mathrm P}
\newcommand{\Fp}{\mathbb F_p}
\newcommand{\LL}{\mathcal L}
\newcommand{\tp}{\operatorname{tp}}
\newcommand{\Sec}{\operatorname{Sec}}
\begin{document}
\maketitle
\pagestyle{empty}
\thispagestyle{empty}
\raggedbottom
\begin{abstract}
We study inflationary fixed-point logic on finite simple graphs with
unordered vertices and an ordered number sort. For a prime \(r\), the
quantifier \(M_r\) tests whether the number of relations satisfying a
formula is nonzero modulo \(r\). For any distinct primes \(p,q\), the
logic with \(M_p\) and \(M_q\) defines the adjacency matrix of a
canonical copy on the number sort. On labelled inputs this matrix is
computable in deterministic polynomial time with a
\(\mathrm{Mod}_p\mathrm P\) oracle. For each prime \(p\), the logic with
\(M_p\) alone cannot define a canonical copy on all graphs. The primes
also differ on Boolean queries. There is a single first-order formula
\(\beta(S)\) on vertex sets such that \(M_q S\,\beta(S)\) is undefinable
with \(M_p\) alone whenever \(p\ne q\). For a positive integer-valued
function \(g\), consider the class of all graphs \(G\) with at most
\(g(|V(G)|)\) automorphisms. For each fixed prime \(p\), the logic with
\(M_p\) canonises this class exactly when \(g\) is bounded.
\end{abstract}

\section{Introduction}

The two endpoints of a single edge are indistinguishable. Neither can
be chosen first by a formula, yet either ordering gives the same adjacency
matrix. For a graph on \(n\) vertices, a \emph{canonisation} specifies an isomorphic graph
on \(I_n=\{0,\ldots,n-1\}\) whose adjacency matrix depends only on the
isomorphism type of the input.

On ordered structures, fixed-point logic captures polynomial time
(PTIME) \citep{Immerman,Vardi}. A definable canonisation lets us apply
this theorem to an unordered graph by evaluating a sentence on its
ordered copy. This is how \citet{Grohe} obtains a logic for PTIME on
graph classes with an excluded minor. On arbitrary graphs, the question
is whether there is a logic with effective syntax that expresses every
PTIME graph property and effectively assigns a polynomial-time
evaluation algorithm to each sentence
\citep{Gurevich}. For a logic that evaluates ordered fixed-point
sentences on a canonical copy, it would suffice to compute that copy in
polynomial time.

The input has an unordered vertex sort and the ordered number sort \(I_n\).
For a prime \(p\), the formula \(M_p R\,\varphi\) holds when the number
of relations \(R\) satisfying \(\varphi\) is nonzero modulo \(p\).
Each relation variable has a fixed arity and specified argument sorts. For example, it may
count vertex sets or binary relations that encode maps from vertices
to numbers. Let \(\LL_p\) be first-order logic with inflationary fixed
points and \(M_p\), and let \(\LL_{p,q}\) allow both modular quantifiers.
Our first theorem separates these languages using only a set of vertices
as the counted relation.

\begin{namedtheorem}[Boolean separation]\hypertarget{th:boolean}{}
There is a first-order graph formula \(\beta(S)\) with one unary relation
parameter such that, for every pair of distinct primes \(p,q\), the graph
property \(M_q S\,\beta(S)\) is undefinable in \(\LL_p\).
Consequently the single-prime languages are pairwise incomparable on
graphs, and adjoining any second prime strictly increases expressive power.
\end{namedtheorem}

To construct \(\beta\), take a connected network over the field
\(\mathbb F_p\) with a prescribed net outflow at each vertex. If the
prescribed values sum to zero, any assignment to the \(d\) edges outside
a spanning tree extends to a unique flow. There are then \(p^d\) flows.
If the total is nonzero, there are none. Counting modulo any other prime
distinguishes these cases. The graph encodes the local constraints, and
each flow corresponds to one set satisfying \(\beta\).

To prove the separation, we must rule out every \(\LL_p\) sentence
that might distinguish the two loads. We use the prime-field version
of the Cai--F\"urer--Immerman construction \citep{CFI,GGPP}. Its
automorphisms form a \(p\)-group, a group whose order is a power of
\(p\). When this group acts on the relations satisfying a formula,
only relations fixed by the whole group contribute modulo \(p\).
We match those fixed relations across the two structures and prove
that the matching persists under further quantification and at each
stage of a fixed point.

\citet[Definition~2.7 and Corollary~3.6]{KS} separated coprime
second-order divisibility quantifiers on a supplied family of sets.
Here a single first-order formula defines the family on simple graphs,
and the competing logic may count relations of any fixed arity.
Affine systems have also supplied lower bounds for infinitary counting
logic \citep{ABD} and for logics with linear-algebraic operators
\citep{GP,DGP}. In our examples Gaussian elimination over \(\mathbb F_p\)
decides whether a solution exists, although the number of solutions
is divisible by \(p\) on both graphs.

\begin{namedtheorem}[Canonisation]\hypertarget{th:main}{}
For any distinct primes \(p,q\), an \(\LL_{p,q}\) formula
\(\gamma_{p,q}(i,j)\) defines the adjacency matrix of a canonical copy of
every finite simple graph on its number sort. No \(\LL_p\) formula defines
such a copy on all finite simple graphs.
\end{namedtheorem}

For the lower bound, a proposed one-prime formula would give the same
adjacency bit at every numerical position on two nonisomorphic CFI
graphs. For the upper bound, let \(Q\) be a Sylow \(p\)-subgroup of the
permutations of \(I_n\), so its order is the largest power of \(p\)
dividing \(n!\). Group two labellings together if they differ by a
permutation in \(Q\). Each class, called a \emph{frame}, has \(|Q|\)
labellings, and the number of frames is prime to \(p\).
Canonical placement under a permutation group \citep{BL} selects one
adjacency string per frame. The inner quantifier \(M_q\) recognises a
frame with a given prefix because \(q\) does not divide the number of
compatible labellings. The outer quantifier \(M_p\) counts each such
frame once. Starting at the empty prefix, we can always extend by a bit
whose count remains nonzero. The final string is a copy of the graph.

Our companion paper \citep{GT} uses the same count of cosets of a Sylow
subgroup. Its algorithm guesses one representative per coset on a
labelled input. Here we represent each coset by a relation on unordered vertices.
The matrix can be computed in polynomial time with an oracle for modular
circuit satisfiability. This oracle tests whether the number of satisfying
assignments is nonzero modulo \(p\), a complete problem for \(\ModP\).
Evaluating ordered fixed-point sentences with this circuit test on the
copy captures the isomorphism-invariant queries in \(\mathrm P^{\ModP}\).

\paragraph*{When one prime suffices.}
A \emph{base} for a permutation group is a tuple of points whose
pointwise stabiliser contains only the identity. We now consider bases for the vertex action of a Sylow \(p\)-subgroup
of the automorphism group \(\Aut(G)\). For a positive integer-valued function \(g\), let
\(\mathcal G_g=\{G:|\Aut(G)|\le g(|V(G)|)\}\).

\begin{namedtheorem}[One-prime canonisation and capture]\hypertarget{th:base}{}
Fix a prime \(p\) and an integer \(k\ge0\). On graphs where a Sylow
\(p\)-subgroup of \(\Aut(G)\) has a base of at most \(k\) vertices,
\(\LL_p\) defines a canonical ordered copy and captures the invariant
\(\ModP\) Boolean queries.

On \(\mathcal G_g\), each of these conclusions holds if and only if \(g\)
is bounded. One PTIME graph property witnesses failure of capture for
every prime \(p\) and every unbounded \(g\). Both equivalences also hold
when \(g\) bounds the Sylow \(p\)-subgroup order or its minimum base size.
\end{namedtheorem}

The entries of a base of bounded size allow a single relation of fixed arity to
encode an entire orbit of labellings. For the converse, we read the
field as well as the flow equations from the graph. Solubility is then
one PTIME property for all primes. Taking the disjoint union of each graph in a CFI pair with the same long path preserves
both answers and doubles each automorphism group order. We can therefore
meet the bound at any sufficiently large size where \(g\) exceeds these
two orders.

We first encode the flow problem on simple graphs and prove that the
one-prime logic preserves the resulting pairs. We then construct the
ordered copy and show how to evaluate queries on it. The final sections
treat graphs with bounded symmetry, the polynomial-time prefix test on
circulants of bounded odd part, and the counts that can be computed on
odd cycles.

\section{Syntax and ordered output}\hypertarget{sec:syntax}{}

Write $V$ for the vertex domain and $I_n=\{0,\ldots,n-1\}$ for the
number domain, where $n=|V|$. Equality compares elements of the same sort. Adjacency is a relation
on \(V\), and order is a relation on \(I_n\). For other input
vocabularies, all input relations likewise have object arguments.
If \(D\) is the product of the argument sorts of \(R\), then
\[
 M_r R\,\varphi(R,\bar a)
 \quad\Longleftrightarrow\quad
 \bigl|\{S\subseteq D : \varphi(S,\bar a)\}\bigr|\not\equiv 0\pmod r.
\]

Each interpretation of \(R\) gives one expansion of the input structure.
\citet{CG} study the number of satisfying expansions as a counting
function. On strings, \citet{KV} study related
second-order quantifiers over unary relations. Here \(R\) may have
arguments of either sort.

The language is closed under Boolean operations and first-order
quantification of either sort. We use the inflationary semantics of
\citet{GS}. A fixed point with body
$\psi(T,\bar x)$ starts at $T_0=\varnothing$ and successively sets
$T_{h+1}=T_h\cup\{\bar a:\psi(T_h,\bar a)\}$. It returns the first stable
relation. Parameters keep their values during iteration. This convention
allows negation and modular tests in the body without a monotonicity
assumption. Simultaneous tables can be represented by one relation with
a numerical tag. Any finite list of exceptional domain sizes in such an
encoding can be handled by first-order cases.

A canonising formula has two free number variables. Its output is the graph
on $I_n$ with the edges it defines, equipped with the usual order. Its graph
reduct must be isomorphic to the input, and isomorphic inputs must have
identical outputs. The latter condition is automatic for formulas in our
language because an input isomorphism acts as the identity on $I_n$.
The formula gives the edges on numbers directly and need never choose a
bijection from $V$ to $I_n$. We apply the Immerman--Vardi theorem only after defining the input to
an ordered computation on the number sort.

\section{A flow problem on graphs}\hypertarget{sec:flow}{}

Fix a prime \(p\) and a connected ordered cubic graph \(D=(W,F)\), where
\(F\) is its set of undirected edges. The CFI structure over \(D\) encodes
a flow problem in a vocabulary independent of \(p\) \citep[Section~5]{GGPP}.
For each orientation \(e\) of an edge, its domain contains a
\emph{fibre} \(\{(e,x):x\in\Fp\}\). The base order induces an order on
the directed edges. A preorder on the domain compares elements by their
edge and ties the elements of each fibre. A successor relation acts by
\(x\mapsto x+1\) in each fibre, and an inverse relation links
\((e,x)\) to \((e^{-1},-x)\).

A \emph{load} \(\lambda:W\to\Fp\) specifies the net outflows. At each
vertex \(w\), list the three outgoing edges in base order. The ternary
relation on their fibres contains exactly the triples whose coordinates
sum to \(\lambda(w)\). Write \(A(D,\lambda)\) for the resulting
structure. Only the relations belong to the vocabulary. The field coordinates
serve to describe the construction.

Write $x\sim y$ for equivalence in the fibre preorder, $J_0(x,y)$ for
the inverse relation, and $U$ for the CFI ternary relation. Define
$U^*(x,y,z)$ to hold if some $x'\sim x$, $y'\sim y$, $z'\sim z$
satisfy $U(x',y',z')$. It records the ordered triples of fibres incident
at a base vertex. A \emph{section} is a set \(S\) that selects exactly one element of each
fibre. We write \(\Sec(S)\) for the first-order condition that \(S\) is
a section satisfying both constraints. Selection of \(x\) forces
selection of its inverse \(y\) whenever \(J_0(x,y)\) holds. For three
selected elements \(x,y,z\), the incidence condition \(U^*(x,y,z)\)
forces \(U(x,y,z)\).

\begin{namedlemma}[Counting flows]\hypertarget{lem:flow}{}
On $A(D,\lambda)$ the number of sets satisfying $\Sec$ is
zero if $\sum_w\lambda(w)\ne0$ and is
$p^{|F|-|W|+1}$ otherwise.
\end{namedlemma}
\begin{proof}
Orient the undirected edges for the calculation. The selected point in
each forward fibre specifies one field value. The inverse condition
fixes the value in the reverse fibre as its negative. The ternary
condition gives exactly the signed incidence equations prescribing
divergence $\lambda$. Summing them gives the necessary zero-total
condition. Conversely, choose a spanning tree and assign arbitrary
values to the $|F|-|W|+1$ edges outside it. Each leaf equation fixes
the remaining incident tree edge. Stripping leaves determines all tree
edges, and the last equation holds by the zero-total condition. Each
resulting flow gives a different set $S$.
\end{proof}

For example, take the complete graph on four vertices as the base over
\(\mathbb F_2\). Three of its six edges lie outside a spanning tree.
Eight sets satisfy \(\Sec\) at total load zero, and none do at total load one.
Counting sections modulo three distinguishes the loads. To rule out
\emph{every} formula with modulus two we will need larger base graphs.

\subsection{Encoding the witnesses as vertex sets}

The graph encoding must preserve the number of witnesses. It must also
let us recognise relations on graph tuples when proving the lower bound.
We use a separate vertex for each tuple and give every auxiliary vertex
a unique code over the original structure.

\begin{namedlemma}[An incidence encoding]\hypertarget{lem:encoding}{}
For a fixed finite relational vocabulary of positive arities there is a
faithful encoding $J(A)$ as a simple graph. Its size depends only on
$|A|$. Each graph vertex has a unique representation by a finite tag and
a bounded tuple of elements of $A$. Adjacency on these codes is
first-order definable. The object domain and its relations are
first-order recoverable in the graph.
\end{namedlemma}
\begin{proof}
Start with an element vertex for each object. For each symbol $U$ of
arity $s$ and \emph{each} $\bar a\in A^s$, add a tuple vertex, a marker,
and $s$ position vertices. Join each position to the tuple and to its
element $a_j$. Join the marker to the tuple precisely when $U(\bar a)$
holds. Distinguish element, tuple, marker and position types, including
the relation symbol and the position index, by attaching a different
constant number of private leaves to each type, always at least two.
The element, tuple, marker and position vertices now have degree at
least two. The private leaves have degree one. The type of each
nonleaf vertex is read from its number of
leaf neighbours. Relations on element vertices are thus recovered by a
fixed first-order formula following the position edges and testing the
marker edge. Every structure isomorphism extends, and every graph
isomorphism induces a structure isomorphism. A private leaf has the code
of its parent and a fixed leaf index. All code domains depend only on
$|A|$, even when membership of tuples changes.
\end{proof}

On simple graphs, define $\beta(S)$ by interpreting $\Sec(S)$ on the decoded
element vertices and requiring every member of $S$ to be an element
vertex. Each flow then gives exactly one witness on the graph. Neither the decoder nor $\beta$
mentions the field size.

\section{Preservation at one prime}\hypertarget{sec:lower}{}

\hyperlink{lem:flow}{The flow count} shows why counting sections modulo \(p\) loses the
total load. To prove undefinability we compare the two CFI structures
on arbitrary \(\LL_p\) formulas. We first describe how a correspondence
between tuple orbits extends to relations.

Fix $K$ and
consider formulas that use at most $K$ variable symbols and allow arbitrary
conjunctions and disjunctions. For each integer $m\ge 0$ they also allow
a quantifier asking whether at least $m$ elements satisfy a formula. The \emph{counting
type} $\tp_K(A,\bar a)$ is the set of these formulas true of $\bar a$ in
$A$. We call two tuples \emph{matched} if their types agree. Numeric
parameters, when present, must agree coordinate by coordinate. The
counting types are taken in the original object vocabulary. We use these types to compare structures in the proof.

\subsection{Matching relations across two structures}

Let $A,B$ be finite structures of the same vocabulary and cardinality, and
let $\Gamma_A,\Gamma_B$ be $p$-subgroups of their automorphism groups.
Suppose that the empty tuples are matched. Up to a tuple length $d\le K$,
suppose also that within either structure two tuples have the same
$K$-type exactly when they lie in the same orbit of its designated group.
Call the tuple of assigned object parameters the \emph{context}. Write
$\Gamma_{A,\bar a}$ for its pointwise stabiliser, the subgroup fixing
each entry of $\bar a$, and define $\Gamma_{B,\bar b}$ likewise. Suppose
the contexts $\bar a,\bar b$ match.
For $|\bar a|+s\le d$, the correspondence
\[
 \Gamma_{A,\bar a}\bar u\ \longleftrightarrow \Gamma_{B,\bar b}\bar v
 \qquad\text{when}\qquad
 \tp_K(A,\bar a,\bar u)=\tp_K(B,\bar b,\bar v)
\]
is a bijection between stabiliser orbits on $s$-tuples. Indeed,
among tuples realised in the two finite structures any one $K$-type can
be isolated by conjoining formulas that distinguish it from the other
realised types. Existentially binding the $s$ new variables gives a test
of realisation over the context, within the same $K$ slots. The tests
agree on matched contexts. Within each side, equality of the extended
types gives a group element sending one extension to the other and fixing
the common prefix. For mixed-sort tuples we
use the same bijection separately at each numerical tuple.

For matched contexts $\bar a,\bar b$, write $\tau_{\bar a,\bar b}(S)$
for the union of the orbits corresponding to those in $S$, whenever $S$
is a union of $\Gamma_{A,\bar a}$-orbits. This transport is a bijection
on invariant relations of each allowed sort and arity. It preserves
unions and the empty relation. In a purely numerical argument domain the
orbits are singletons and transport is the identity.

The formula determines how long a tuple we must match. Let \(t\) be the
number of object variables in the initial context, including all free
object variables. Give distinct names to all bound variables. Let \(b\)
be the total number of object variables bound by individual quantifiers
or introduced as arguments of fixed-point bodies. Let \(r\) be the
largest number of object arguments of any relation variable in the
formula, taking \(r=0\) if there are none. A bound \(d\ge t+b+r\)
then covers the context of every subformula together with one argument
tuple for a relation. Repeated fixed-point stages use the same variables.

\begin{namedlemma}[Preservation under modular relation quantification]\hypertarget{lem:preserve}{}
Let $\varphi\in\LL_p$ and let $A,B$ satisfy the orbit and counting-type
hypotheses above through length $d\ge t+b+r$. Evaluate $\varphi$ at
matched contexts $\bar a,\bar b$. For each free relation let its value
$S$ on $A$ be invariant under $\Gamma_{A,\bar a}$ and give it value
$\tau_{\bar a,\bar b}(S)$ on $B$. Then the two evaluations agree.
\end{namedlemma}
\begin{proof}
We argue by induction on the formula, retaining the outer context at each step.

\emph{Extending the context.} If
$(\bar a,\bar u)$ and $(\bar b,\bar v)$ match, every relation $S$ invariant
at $\bar a$ is invariant at $(\bar a,\bar u)$. Moreover,
\[
 \tau_{\bar a\bar u,\bar b\bar v}(S)=\tau_{\bar a,\bar b}(S).
\]
Corresponding orbits in the extended contexts have the same extended
type. Forgetting the new coordinates sends them to corresponding
orbits in the original contexts. Membership in \(S\) is constant on
each original orbit, which proves the equality. In particular, a relation
bound by an outer modular quantifier keeps the same interpretation when
an inner quantifier introduces a new parameter.

Atomic formulas agree by matched types, equality of numerical arguments,
and the assumption on free relations. The Boolean steps follow
immediately. For a first-order existential formula, extend the context
by a witness. The orbit correspondence supplies a matched extension on
the other side. For a numerical witness use the same number. The
induction hypothesis applies, and the argument is symmetric.

\emph{Binding a relation.} For $M_p R\,\psi$ in $A$, the stabiliser of $\bar a$ acts on
the set of all assignments to $R$ that satisfy $\psi$. This set is closed
under the action because the other free relations are invariant and
formula evaluation commutes with automorphisms for every assignment to $R$.
An automorphism carries the witnesses of each inner modular test bijectively
to those for the transported parameters, and it carries each fixed-point
stage to the corresponding stage. Each orbit with more than one assignment has size divisible by \(p\).
It follows that the count modulo \(p\) depends only on the satisfying
assignments fixed by the stabiliser. Each fixed assignment to \(R\) is a
union of orbits on its argument tuples. The orbit correspondence pairs
these unions bijectively across the two structures. Applying the induction
hypothesis to \(\psi\) with each corresponding pair of assignments shows
that equal numbers of them satisfy \(\psi\). The two modular counts
therefore agree.

\emph{Iterating a fixed point.} Let $F_A(T)$ be one inflationary update in the original context
$\bar a$, and define $F_B$ likewise. Write $\tau=\tau_{\bar a,\bar b}$.
If the original stabiliser fixes $T$, it also fixes $F_A(T)$, since
formula evaluation commutes with automorphisms. At matched argument tuples $\bar u,\bar v$,
the refinement identity allows us to pass $T,\tau(T)$ and all outer
relations to the body. The induction hypothesis for the body gives the
same decision about adding the tuple on both sides. Old membership also
agrees. Hence the updates satisfy
\[
                 \tau(F_A(T))=F_B(\tau(T)).
\]
Starting from the empty relations gives $T_h^B=\tau(T_h^A)$ at every
stage $h$. Bijectivity of $\tau$ implies that a stage changes on one side
exactly when it changes on the other. At their common stable stage,
membership agrees at the argument of the fixed-point formula. This stage argument uses the induction hypothesis only for the update
body, which may itself contain fixed points and modular quantifiers.
\end{proof}

\subsection{Applying the lemma to the encoded graphs}

On a common connected base, two CFI structures are isomorphic exactly
when their total loads agree. Their automorphisms act by translations
on the edge fibres and form a \(p\)-group
\citep[Section~5.2 and Theorem~5.2]{GGPP}.

Two further results apply on the ordered cubic expander bases.
For every fixed \(K\), structures with different total loads on a
sufficiently large common base agree on all \(K\)-variable counting
sentences \citep[Theorem~5.6]{GGPP}. Within either structure, the
counting type in \(\ell s\) variables determines the automorphism orbit
of an \(s\)-tuple, for a constant \(\ell\) independent of the base
\citep[Theorem~5.7]{GGPP}. Given the tuple bound \(d\), choose
\(K\ge\ell d\) large enough to read the vocabulary and then choose the
base. This gives both hypotheses of the
\hyperlink{lem:preserve}{preservation lemma} on the CFI structures.

\begin{proof}[One-prime lower bound]
Fix a proposed canonising formula. Multiply its tuple-length bound by
the maximum number of object coordinates in a graph vertex code
(\hyperlink{lem:encoding}{the incidence encoding}). Choose CFI structures \(A,B\) on the same
base with total loads zero and one, using this enlarged bound. Let
\(\Aut(A)\) act on \(J(A)\) through the coordinates of the codes, fixing
each tag, and use the corresponding action on \(J(B)\). These designated
groups are \(p\)-groups even though private leaves with the same parent can also be permuted.
The tags specify this action on the constructed graphs. No tag or leaf
index is supplied to the proposed canonising formula.

Match two graph tuples when their tags agree and their concatenated
object coordinates have the same counting type. For fixed tags,
extending a graph tuple amounts to adjoining a bounded number of
object coordinates. The uniqueness of each code therefore gives the
orbit correspondence on graph tuples. Forgetting a
graph parameter removes exactly its tag and object coordinates, so the
refinement identity for invariant relations also holds. Atomic adjacency
agrees because it is first-order definable on codes, with \(K\) chosen
large enough for the decoding formula. The graphs have the same size,
so their numerical parameters match by equality.

The induction in \hyperlink{lem:preserve}{the preservation lemma} uses
only agreement on atomic formulas and a bijection of extension orbits
compatible with forgetting coordinates. The matching on graph codes
supplies both, so the proposed formula gives the same answer on the
two graphs at each numerical pair \((i,j)\). A single matrix cannot
be a copy of two nonisomorphic graphs.
\end{proof}

\begin{proof}[Boolean separation]
Fix $q$ and any prime $p\ne q$. On graphs encoding CFI structures over
$\Fp$, \hyperlink{lem:flow}{the flow count} gives either zero witnesses or a power of
$p$. Consequently $M_q S\,\beta(S)$ accepts exactly the zero-total
loads on these encodings. If an $\LL_p$ sentence defined this graph
query, the preceding argument with empty context would give it the same truth
value on the two encoded structures. This contradicts the count.
Exchanging the roles of $p$ and $q$ gives incomparability. Syntactic
inclusion gives $\LL_p\subseteq\LL_{p,q}$ and the exhibited sentence
makes this inclusion strict.
\end{proof}

\section{A canonical matrix with two primes}\hypertarget{sec:upper}{}

Fix distinct primes \(p,q\). A frame groups together the labellings
that differ by a permutation in a fixed Sylow \(p\)-subgroup of \(S_n\).
To count frames with a relation quantifier, we need one relation per
class of labellings. We construct that relation from the colours of
ordered pairs of positions.

\subsection{The frame and its labellings}

Write $n=\sum_h a_hp^h$ in base $p$. Divide $I_n$ into consecutive slots
in decreasing order of size, with $a_h$ separately indexed slots of size
$p^h$. Identify the members of such a slot with the words of length $h$
over $\Fp$ in digit order. For a pair of positions record both slot indices.
For distinct words in the same slot record also the first position at which
they differ and the difference between their digits there, taken in
argument order (second minus first).
Use separate tags for a diagonal pair and a pair in different slots, and
fix every unused coordinate to zero. Regard this table as a relation
$\mathsf{Col}_n(i,j,\bar c)$ whose unique value $\bar c$ is a fixed-length numerical
tuple. For fixed $p$ its entries are uniformly computable in polynomial
time, so an ordered fixed-point formula defines $\mathsf{Col}_n$. Write $v_p(n!)$
for the exponent of $p$ in $n!$.

\begin{namedlemma}[The group of a frame]\hypertarget{lem:sylow}{}
The automorphism group $Q$ of the table on its position domain, with
colours fixed, has order $p^{v_p(n!)}$.
\end{namedlemma}
\begin{proof}
Each slot is preserved. Within a slot the first differing position tells
which pairs agree on any given prefix. A table automorphism therefore acts
on the rooted tree of prefixes. If it sends a node to another node, the
induced bijection $\sigma$ of their child digits satisfies
$\sigma(b)-\sigma(a)=b-a$ for all $a,b$. It is a translation. Conversely,
choosing a translation at each internal node defines a tree automorphism
that preserves the table. A slot of size $p^h$ has $(p^h-1)/(p-1)$
internal nodes. Thus the exponent of $p$ in $|Q|$ is
\[
 \sum_h a_h\frac{p^h-1}{p-1}
 =\sum_{j\ge1}\left\lfloor n/p^j\right\rfloor=v_p(n!).
\]
\end{proof}

For a bijection $f:V\to I_n$ let $H_f(u,v,\bar c)$ mean
$\mathsf{Col}_n(f(u),f(v),\bar c)$. Two maps $f,g$ give the same relation exactly
when $g f^{-1}\in Q$. Each frame occurs as the pullback of exactly $|Q|$
maps. Consequently there are $n!/|Q|$ frames, a number nonzero modulo $p$.

\subsection{Reading a frame and choosing the prefix}

The cyclic shifts of the children at individual nodes generate \(Q\).
Each shift leaves suffix digits unchanged, and there are only polynomially
many nodes.
Acting on unordered pairs of positions, these generators again give
a \(p\)-group, hence a solvable group. The canonical placement algorithm
of \citet[Theorems~3.3 and~3.5]{BL} therefore computes in
polynomial time a string \(\CF_Q(X)\) in the orbit of \(X\), with the
same output for all strings in that orbit. We order pairs \(i<j\)
lexicographically, use \(0<1\) for bits and follow their first-orbit
and first-minimal-block-system rules (Section~3.2).
For any labelling \(f\) compatible with a frame \(H\), put
\[
                      D(G,H)=\CF_Q(f(G)).
\]
All compatible labellings give the same string because they differ by
elements of \(Q\).

Encode a prefix $s$ by its length or endpoint and its known bits in the
lexicographic order on pairs $i<j$. Define $\theta(G,H,f,s)$ by requiring
that $f\subseteq V\times I_n$ be the graph of a bijection, that
$H(u,v,\bar c)$ agree with $\mathsf{Col}_n(f(u),f(v),\bar c)$ on every argument,
and that $\CF_Q(f(G))$ start with $s$. The last test is a polynomial-time
predicate on an ordered graph with numerical parameters encoding $s$.
Translate it to ordered fixed-point logic and substitute
\[
 \exists u\,\exists v\,
       (f(u,i)\wedge f(v,j)\wedge E(u,v))
\]
for the transported edge relation. The resulting formula for $\theta$ uses only first-order quantifiers
and fixed points.

\begin{namedlemma}[Counting frames with a given prefix]\hypertarget{lem:prefix}{}
The formula
\[
 M_p H\, (M_q f\,\theta(G,H,f,s))
\]
holds exactly when the
number of frames $H$ for which $D(G,H)$ begins with $s$ is nonzero modulo $p$.
\end{namedlemma}
\begin{proof}
For an invalid table no bijection passes the pullback test. For a valid
frame the number of maps passing all three tests is either zero or $|Q|$.
Since $|Q|$ is a power of $p$, the inner quantifier accepts exactly the
valid frames with the specified prefix. Each such frame contributes one
assignment to the outer quantifier.
\end{proof}

Take a graph consisting of two disjoint edges and let \(p=2\).
The three frames are the three partitions into two pairs, each with eight
compatible labellings. One frame pairs the endpoints of both edges and
places a one first. The other two pair non-neighbours and place a zero
first. With \(q=3\) the inner test recognises each eligible frame since
eight is nonzero modulo three. Two frames extend the empty prefix by zero and one frame extends it
by one. The parity test therefore chooses one. Counting labellings by
parity would give zero for both children.

\begin{proof}[Two-prime construction]
Let $c_G(s)$ count the frames whose placed string begins with $s$. The count at the empty prefix
is prime to $p$. If the current count is nonzero modulo $p$, then
$c_G(s0)+c_G(s1)=c_G(s)$ implies that at least one child count is nonzero
modulo $p$. Choose zero when the test in \hyperlink{lem:prefix}{the prefix-counting lemma} accepts
$s0$, and one otherwise. An inflationary fixed point stores triples
$(i,j,b)$ for the bits already decided. Its body adds the value for the
first undecided pair using the current table and the prefix test. By
induction every stage is a prefix with nonzero count. After $\binom n2$
steps some frame realises the entire string. That string is a labelled
copy of the input. Relabelling the input transports frames bijectively
and preserves their strings, so it preserves all the decisions. Read
the stored bits symmetrically, with zero diagonal. For $n\le1$ the output
edge relation is empty.

\end{proof}

\paragraph*{Computing the same matrix with one oracle.}
On a labelled input, guess one membership bit for each entry of the frame
table in a fixed order. Each table occurs on exactly one branch.
The following procedure recovers a compatible map from every valid table.
Read slot indices from diagonal entries. Within each slot group leaves
by agreement through each prefix length, using the first-difference
field. Check that these are nested partitions of the required sizes.
At each internal node take the child containing the least input vertex
for digit zero and use the recorded
differences from it to place the others. Reject if any test fails. Once all digits have been assigned, compare
the entire table with the pullback along the resulting bijection.
This equality certifies that the table is a frame. On a valid frame,
translation symmetry permits any choice of the zero child.
Compute $\CF_Q$ using this map and check the prefix. The accepting path
count is exactly $c_G(s)$. A $\ModP$ oracle answers the zero-child query
at each of polynomially many positions.

\section{Evaluation on the ordered copy}\hypertarget{sec:evaluation}{}

For any polynomial-time graph property $P$, the Immerman--Vardi theorem
gives an IFP sentence $\psi(E,<)$ defining $P$ on ordered graphs. Run $\psi$ on the
number sort, replacing each adjacency test $E(i,j)$ by
$\gamma_{p,q}(i,j)$. By \hyperlink{th:main}{the canonisation theorem} the graph on which $\psi$
runs is isomorphic to the input, so the resulting sentence defines $P$.
Once the matrix is known, evaluating $\psi$ takes ordinary polynomial time.

We can also characterise the queries computable with the same oracle
that constructs the matrix. Keep $\gamma_{p,q}$ fixed and allow ordered
IFP sentences to test whether a definable Boolean circuit has a nonzero
number of satisfying assignments modulo $p$. Index the gates by
fixed-length number tuples and reject circuit descriptions that are
malformed or cyclic. To express this test, bind a relation of gate values by
$M_p$, check every gate and the output, and force unused tuple entries
to be zero. Each satisfying assignment to
the input gates has a unique extension to all gates.

\begin{namedcorollary}[Evaluation on the ordered copy]\hypertarget{cor:ordered}{}
Ordered fixed-point logic with modular circuit tests, evaluated on the copy defined by \(\gamma_{p,q}\), captures the isomorphism-invariant
Boolean graph queries in $\mathrm P^{\ModP}$.
\end{namedcorollary}
\begin{proof}
First compute the canonical copy with a $\ModP$ oracle. For any fixed
ordered sentence the relation tables and fixed-point iterations are
polynomially bounded. Each circuit call has polynomial size and takes one
oracle query after its description has been evaluated. Fixed nesting of
subformulas preserves the polynomial bound.

For the converse, take modular circuit satisfiability as the complete oracle
problem. A polynomial-time nondeterministic verifier can be represented
by a circuit on its polynomially many choice bits with unused bits fixed,
preserving the path count. Fix a deterministic polynomial-time oracle
machine for the query. On the ordered copy, tuples index time, tape cells
and query gates up to the machine's polynomial bound. An inflationary
table records the configuration at each time. Its successor row is given
by the fixed transition rule, using the circuit test on the preceding
query tape at an oracle instruction. The initial row and the decoding
of the query tape are polynomial-time ordered predicates and hence are
definable by ordered fixed points. The accepting state defines the desired
sentence. Isomorphism invariance gives the correct answer on the canonical
copy.
\end{proof}

\section{One prime under symmetry bounds}\hypertarget{sec:base}{}

The \hyperlink{sec:upper}{two-prime construction} counts a frame by testing
its labellings with a second prime. Under the base hypothesis of
\hyperlink{th:base}{the one-prime canonisation theorem}, we can count orbits of labellings directly.
\citet[Theorem~5.12]{Wielandt} showed that the orbits on
\((k+1)\)-tuples determine a permutation group with a base of size \(k\).
\citet{KoV} used bases for the full automorphism group to reduce
parallel canonisation to a complete invariant for coloured graphs.
Here the base bound applies to a Sylow subgroup. The input supplies
neither the subgroup nor a base. We perform the group computations on
the number sort once a candidate graph has been specified.

When \(k=0\), we have \(p\nmid|\Aut(G)|\), and exactly \(|\Aut(G)|\) bijections onto \(I_n\) give
any specified copy. On the number sort a one-prime formula can test
whether that copy equals the canonical output, using the
\hyperlink{sec:evaluation}{ordered simulation of the oracle machine}. Counting the bijections that
pass this test and give a one at \((i,j)\) recovers the canonical bit.
For larger \(k\) the same multiplicity can be zero modulo \(p\).
We therefore count orbits of bijections under a suitable subgroup.
A related result of \citet[Lemma~3.10]{FJ} says that two graphs
without automorphisms of order \(p\) are isomorphic if their
homomorphism counts from every graph agree modulo \(p\).

\subsection{Encoding an orbit of labellings by one relation}

Keep the Sylow subgroup $Q\le S_n$ and the canonical placement function
$\CF_Q$ from the \hyperlink{sec:upper}{two-prime construction}. For an ordered candidate graph $Y$
on $I_n$, put $H_Y=Q\cap\Aut(Y)$, where automorphisms here preserve the
graph reduct of $Y$. 
We can compute generators of $H_Y$ in polynomial time on the number
sort by the Babai--Luks stabiliser algorithm for the solvable group
$Q$ acting on pair positions \citep[Lemma~3.2 and Theorem~3.5]{BL}. For $n\ge3$ this
action is faithful, and a permutation of vertices is recovered from
its action on incident pairs. Smaller graphs can be handled directly.

Suppose \(G\) satisfies the base bound \(k\) of \hyperlink{th:base}{the one-prime canonisation theorem}.
If \(Y\) is a copy of \(G\), then \(H_Y\) lies in a Sylow \(p\)-subgroup
of \(\Aut(Y)\). By Sylow conjugacy that subgroup has a base of size at
most \(k\), which is also a base for \(H_Y\). We need only a base for
this last group. For example, on a triangle a single vertex can be a
base for a Sylow \(2\)-subgroup even though its stabiliser in the full
group \(S_3\) contains a transposition.

Call a candidate \(Y\) \emph{admissible} if \(\CF_Q(Y)=Y\) and \(H_Y\)
admits a base in \(I_n^k\), allowing repetitions. For an admissible
candidate take the lexicographically first such tuple \(\bar b\).
Compute \(|H_Y|\) from its generators \citep{FHL}. For fixed \(k\)
each tuple orbit has size at most \(n^k\), so we can enumerate it in
polynomial time and test whether its size equals \(|H_Y|\). At \(k=0\) the sole candidate base is the
empty tuple, which passes precisely when \(H_Y\) is trivial.
Orbit membership for \((k+1)\)-tuples is polynomial-time decidable by
the same enumeration. Ordered fixed-point logic defines both tests.

For an isomorphism $f:G\to Y$ define the mixed-sort relation
\[
 R_f(\bar x,u,i)\ \Longleftrightarrow\
       (\bar b,i)\in H_Y\cdot(f(\bar x),f(u)),
 \qquad \bar x\in V^k,\quad u\in V,\quad i\in I_n .
\]
Its slice at $\bar x$ is empty unless some $h\in H_Y$ sends
$f(\bar x)$ to $\bar b$. Such an $h$ is unique because $\bar b$ is a
base. The slice is then the graph of $hf$. Conversely, every $hf$
occurs in the slice at $(hf)^{-1}(\bar b)$. Thus the distinct nonempty
slices are precisely the maps in the orbit $H_Y f$.

\begin{namedlemma}[One relation per orbit of labellings]\hypertarget{lem:orbitcode}{}
For each admissible $Y$, a formula with first-order quantifiers on the
vertices and fixed points on the number sort recognises exactly the
relations $R_f$. There is one such relation per $H_Y$-orbit of
isomorphisms from $G$ to $Y$.
\end{namedlemma}
\begin{proof}
For a proposed relation $R$, existentially quantify a vertex tuple
$\bar x_0$ and read the slice $R(\bar x_0,u,i)$ as the graph of a map
$g$. Verify that it is a bijection carrying the edges and nonedges of
$G$ to those of $Y$. Test the displayed equivalence defining $R_g$ on
every argument, using that slice wherever $g$ occurs. If the test succeeds, $R=R_g$. For $R=R_f$ the slice at
$\bar x_0=f^{-1}(\bar b)$ gives $g=f$, so the test succeeds.
Postcomposition by $H_Y$ preserves $R_f$. Conversely, $R_f=R_{f'}$ implies that the slice equal to
$f'$ also occurs in $R_f$, so $f'=hf$ for some $h\in H_Y$.
The tuple $\bar x_0$ is an ordinary existential witness
inside the test, so choosing it introduces no multiplicity into the count
of relations.
\end{proof}

For a prefix $s$, count pairs $(Y,R)$ for which $Y$ is an admissible
$Q$-canonical graph extending $s$ and $R$ satisfies the \hyperlink{lem:orbitcode}{orbit encoding conditions}.
A single binder $M_p$ counts these pairs. Reserve two numerical tags
in one mixed-sort relation, one for $Y$ and one for $R$. For each tag,
require membership to be independent of the unused coordinates. Forbid
entries with any other tag. This encodes each pair exactly once on
domains of size at least two. The candidate adjacency relation is
required to be symmetric with empty diagonal.

Let \(a=|\Aut(G)|\). There are \(a\) isomorphisms from \(G\) to a
specified copy \(Y\). The free action of \(H_Y\) on these maps gives
\(a/|H_Y|\) distinct relations for \(Y\). The \(Q\)-orbit of \(Y\), in
turn, contains \(|Q|/|H_Y|\) labelled graphs. Sum over the unique
\(Q\)-canonical representative of each orbit of copies. All these
representatives are admissible, so the count at the empty prefix is
\[
   \sum_Y\frac{a}{|H_Y|}
   =\frac{a}{|Q|}\sum_Y|Q\cdot Y|
   =\frac{a}{|Q|}\frac{n!}{a}
   =[S_n:Q].
\]
This integer is prime to \(p\). The same pairs \((Y,R)\) partition at
each prefix according to the next bit of \(Y\). We can therefore use
the \hyperlink{sec:upper}{prefix fixed point} to choose a full string.
Relabelling \(G\) transports \(R\) and fixes \(Y\), so every input
labelling gives the same string.

On labelled inputs the data complexity of every fixed $\LL_p$ formula
is in $\ModP$. To see this, induct on syntax with the current free
relations supplied as explicit tables. A modular binder guesses one
table and tests its body by an oracle call. Individual quantification
and fixed-point iteration need only polynomially many such calls for
a fixed formula. The induction closes by the prime-modulus lowness
theorem $\ModP^{\ModP}=\ModP$ \citep[Theorem~6.7 and Corollary~6.8]{BG}.
Conversely, on the promised class the matrix formula supplies an ordered
copy. The \hyperlink{sec:evaluation}{ordered circuit test} then expresses
every invariant $\ModP$ query with the same prime. Thus the sentences of
$\LL_p$ define exactly the restrictions of those queries to the class.
A polynomially clocked machine correct only on the promise gives the
same conclusion. Its simulation on the definable matrix is invariant
on all inputs and agrees with the query on the promised class, even
when membership in the class is undecidable.

\subsection{One polynomial-time witness for all primes}

We define a Boolean query $U$ on all simple graphs. First apply the
first-order decoder of \hyperlink{lem:encoding}{the incidence encoding} for the CFI signature.
Require a nonempty decoded domain, a total preorder on it, and a
successor relation consisting of one directed cycle on each equivalence
class of the preorder. Require all cycles to have the same prime length
$r$. Reject if any of these tests fails.

Choose an origin in each cycle and follow its successor to give its
elements coordinates in $\mathbb F_r$. Introduce one unknown $z_C$
in that field per class $C$. For the decoded inverse and ternary
relations, partition their tuples by their ordered lists of classes.
Each nonempty block with list $(C_1,\ldots,C_j)$ gives a table
$B\subseteq\mathbb F_r^j$, where $j=2$ or $3$. Require $B$ to be an
affine subspace, and impose the constraint
$(z_{C_1},\ldots,z_{C_j})\in B$. Empty blocks contribute no constraint.
Accept exactly when all tests pass and the resulting system is soluble.

This defines an invariant polynomial-time query even on malformed
encodings. Gaussian elimination computes the affine hull of each
nonempty block and equations for it. A block with hull of dimension $e$
equals its hull exactly when it has $r^e$ entries. Solve the combined
system by elimination. The field size is at most the input size.
Changing the origins translates each unknown and its occurrences in all
tables by the same offset. Thus solubility and the tests for affine
subspaces are independent of these choices. On an encoded
$A(D,\lambda)$ the inverse blocks impose negation and the ternary
blocks prescribe the sum at each vertex. \hyperlink{lem:flow}{The flow count} gives
\[
            U(J(A(D,\lambda)))\quad\Longleftrightarrow\quad
                           \sum_w\lambda(w)=0 .
\]
The field is read from the graph, so the query is common to all primes.

\citet{BGS} used padded CFI structures to separate fixed-point logic
with counting from choiceless polynomial time. Their padding by isolated vertices permits arbitrary
permutations of those vertices. Here the size of the full automorphism
group is part of the hypothesis, so we use a single path.

Fix \(p\), an unbounded \(g\), and a proposed \(\LL_p\) formula. Choose
its encoded CFI pair \(X_0,X_1\) as in the \hyperlink{sec:lower}{one-prime lower bound}.
Let \(m\) be their common order and \(a_*=\max_i|\Aut(X_i)|\).
Every tail of \(g\) is unbounded, so there exists \(N\) with
\(N-m>\max(m,3)\) and \(g(N)\ge2a_*\). Add a disjoint path on
\(N-m\) vertices to each graph. It is larger than every old component
and has two automorphisms. The new automorphism group order is therefore
\(2|\Aut(X_i)|\), and both padded graphs lie in \(\mathcal G_g\).
No path vertex has the two leaf neighbours required by the decoder,
so \(U\) still distinguishes the graphs.

Extend the designated \(p\)-group on each core \(X_i\) by the identity
on the path. Match core coordinates as in the \hyperlink{sec:lower}{one-prime lower bound} and
match path vertices and numbers by equality. Once the path vertices
and numbers of a tuple are specified, its stabiliser orbit is determined
by its core coordinates. The orbit correspondence therefore extends to
mixed tuples and still commutes with forgetting parameters. Adjacency
is preserved, since the path and the core are disjoint.

These are the hypotheses used in the induction of
\hyperlink{lem:preserve}{the preservation lemma}. Modular binders match unions of the extended
orbits, and fixed-point updates preserve the matching at every stage.
The formula's tuple-length bound is independent of \(N\). Thus the two
padded graphs agree on the proposed sentence, although \(U\) distinguishes
them. A proposed canoniser likewise assigns the same matrix to the two
nonisomorphic graphs.

For the bounded case, suppose \(|\Aut(G)|\le C\). Whenever the current
pointwise stabiliser in a Sylow \(p\)-subgroup is nontrivial, choose a
vertex that it moves. Fixing that vertex decreases its order by a factor
of at least \(p\). After at most \(\lfloor\log_p C\rfloor\) choices the
stabiliser is trivial, so the bounded-base result applies. The same
argument works when \(C\) bounds just the Sylow order, and a bound on
the base size is the hypothesis itself. Conversely, the padded examples
meet all three unbounded bounds because the full group order bounds
both other parameters. The equivalences in \hyperlink{th:base}{the one-prime canonisation theorem}
concern classes specified by a numerical bound alone. Particular graph
classes with unbounded automorphism groups, such as the complete
graphs, can still admit a definable canonical copy.

\section{What remains for polynomial-time evaluation}\hypertarget{sec:ptime}{}

The construction expresses every PTIME graph property on the ordered
copy. Its evaluation cost comes from the modular prefix test. \citet{LS} obtain a different connection in choiceless polynomial time
with witnessed symmetric choice. In their logic, definable isomorphism
yields definable canonisation on classes closed under individualisation.

For the full language, modular circuit satisfiability already occurs on
simple graphs. Encode the gate types and wires of a Boolean circuit by
\hyperlink{lem:encoding}{the incidence encoding}. A first-order formula requires a vertex set to
consist exactly of the gates assigned true in a consistent evaluation
with true output. Each satisfying input has one such extension, so \(M_r\)
applied to this formula tests modular circuit satisfiability. The encoding
takes polynomial time. A polynomial-time evaluator for every fixed
\(\LL_{p,q}\) sentence would therefore put both \(\mathrm{Mod}_p\mathrm P\)
and \(\mathrm{Mod}_q\mathrm P\) in \(\mathrm P\).

\paragraph*{The prefix test.}
Fix the primes \(p,q\) and the \hyperlink{sec:upper}{canonical placement algorithm}. Call \(s\) \emph{reached} if it occurs during the construction of the
matrix on \(G\). Consider the promise problem with input \(G\) and a
reached, unfinished prefix \(s\). Is \(c_G(s0)\) nonzero modulo \(p\)?

The test and the matrix are polynomial-time equivalent. Starting from the
empty prefix, \(\binom n2\) calls recover the matrix, and each call uses
the prefix reached so far. Conversely, given the matrix, read the bit
after \(s\). It is zero exactly when \(c_G(s0)\) is nonzero modulo \(p\).
A polynomial-time algorithm for the test would therefore give a logic
capturing PTIME by evaluation on the copy. Comparing the matrices of two
graphs would also decide isomorphism in polynomial time.

\subsection{Circulant graphs of bounded odd part}

A graph is \emph{circulant} if it has an automorphism consisting of a
single cycle through all its vertices. If the odd part of the cycle length is bounded,
only polynomially many frames are fixed by the subgroup of rotations of
power-of-two order.

\begin{namedproposition}[Circulants of bounded odd part]\hypertarget{prop:circulant}{}
Fix \(M\). On circulant graphs of order \(n=m2^h\), with \(m\) odd
and \(m\le M\), every frame-prefix parity is computable in deterministic
polynomial time from the graph alone. Consequently ordinary ordered IFP
evaluated on the copy \(\gamma_{2,q}\), for any odd prime \(q\), captures
PTIME on this class.
\end{namedproposition}
\begin{proof}
\citet[Theorem~1.1]{EP} compute a cycle base in polynomial time.
On a circulant graph this gives a full-cycle automorphism \(\sigma\). Put \(L=2^h\). The permutation
\(\tau=\sigma^m\) has \(m\) cycles of length \(L\). The cyclic group
generated by \(\tau\) has power-of-two order and preserves the placed
string of each frame it moves. Every prefix parity is therefore the
parity of its matching \(\tau\)-fixed frames.

When \(m=1\) there is just one such frame. Number the vertices along
\(\sigma\) in \(\mathbb Z/2^h\mathbb Z\). Each level of a fixed binary
tree is a partition invariant under translations. The block containing zero is a subgroup, and the other blocks are its
cosets. At depth \(j\)
the cyclic group has a unique subgroup of index \(2^j\). Thus the
levels are exactly the partitions into residue classes modulo \(2^j\).
One placement of this explicit frame answers all prefix queries.

For bounded \(m\), the top slots have distinct sizes \(kL\), with \(k\)
ranging over the powers of two occurring in \(m\). Assign whole cycles
of \(\tau\) to these slots in at most \(m!\) ways. Let \(F(k,L)\) count
the balanced trees on one slot preserved by its \(k\) cycles. We have
\(F(1,L)=1\), and \(F(k,1)\) counts all balanced trees on \(k\) leaves.
For \(k,L\ge2\),
\[
 F(k,L)=\tfrac12\binom{k}{k/2}F(k/2,L)^2+2^{k-1}F(k,L/2).
\]
In the first term each root child receives half the cycles. In the
second term \(\tau\) exchanges the children, so their leaves alternate
around each cycle. There are \(2^{k-1}\) unordered choices of halves.
One half carries \(k\) cycles of \(\tau^2\) of length \(L/2\).
Choosing its tree determines the other by transport.

For fixed \(k\), induction on \(k\) gives \(F(k,L)=O_k(L^{k-1})\).
Indeed the first summand is \(O_k(L^{k-2})\). Dividing by \(L^{k-1}\)
gives \(f(L)\le f(L/2)+O_k(1/L)\), where \(f(L)=F(k,L)/L^{k-1}\).
The sum of the error terms over powers of two is bounded.
The same splits enumerate the trees. Always process the child
containing the least input vertex first to avoid duplicates. Every partial choice extends to an output, and the path to an output
has at most \(n-1\) splits. Thus the enumeration visits at most \(n\)
partial choices per output on average. Each choice takes polynomial time.
The bound on \(m\) makes the product of the slot counts polynomial in
\(n\). Run placement on each resulting frame and count the requested
prefix modulo two. The answer depends only on the original frame count,
so any full cycle returned by the algorithm gives the same answers.
The capture assertion follows by evaluating ordered IFP on this copy.
\end{proof}

\section{The parity test on odd cycles}
Numbering an odd cycle consecutively gives it a canonical form. To
compute the matrix defined by the \hyperlink{sec:upper}{two-prime construction}, we must
recover the bits chosen by the parity test. Each frame is a forest of full binary trees, one for each
power of two in the binary expansion of \(n\). The singleton slot
provides a distinguished leaf. Rotate the cycle until this leaf is at
vertex \(0\). Each anchored frame then accounts for \(n\) rotations,
an odd number. Reflection about \(0\) pairs the remaining frames unless
it preserves their trees. Let \(\mathcal F^\rho\) denote the anchored frames fixed by this reflection. For every prefix \(t\),
\[
 c_{C_n}(t)\equiv
 \bigl|\{F\in\mathcal F^\rho : D(C_n,F)\text{ begins with }t\}\bigr|\pmod 2 .
\]

For example, on \(C_7\) there are nine anchored frames fixed by
reflection. Seven place \(00\) first. Of those seven, three continue
with zero and four with one, so the algorithm chooses \(000\).
Counting all labelled copies of the cycle would assign a different
multiplicity to the same strings.

For the counts below, fix \(p=2\) and write \(n=2k+1\ge5\).
Let \(B=2b\) be the largest power of two at most \(n\). Write the cycle in the order
\(0,1,\ldots,k,-k,\ldots,-1\), with singleton slot at \(0\).
Reflection sends \(i\) to \(-i\). We use the placement algorithm and
numerical pair order of the \hyperlink{sec:upper}{two-prime construction}. On \(C_3\) every
placement gives the string \(111\).

\subsection{Counting at the beginning and end of the matrix}

The \(Q\)-orbit of the first output position \((0,1)\) consists of the
\(b\) sibling pairs in the largest slot. The Babai--Luks algorithm
minimises the bit at the least position of its current domain. In an
intransitive call it places the orbit of that position first. In a
transitive call it keeps the least of the recursively placed strings.
Induction on these two cases gives the claim. The first placed bit
is therefore one exactly when all \(b\) sibling pairs are cycle edges.

For a power of two \(h\), let \(T(h)\) count the balanced binary trees
on \(2h\) specified leaves preserved by an involution with \(h\) transpositions.
This is \(F(h,2)\) in the \hyperlink{prop:circulant}{count for circulant graphs}. We have \(T(1)=1\)
and, for \(h\ge2\),
\[
 T(h)=h!+\tfrac12\binom{h}{h/2}T(h/2)^2 .
\]
The factorial counts trees whose two halves are exchanged. The half
containing the least leaf contains one leaf from each transposition,
giving \(2^{h-1}\) choices for that half. Its leaves admit
\(h!/2^{h-1}\) trees, each of which determines the reflected tree in the
other half. The second term counts the trees whose halves are preserved.

Write \(r=\sum_j h_j\) as a sum of distinct powers of two. There are
\[
 A(r)=r!\prod_j\frac{T(h_j)}{h_j!}
\]
ways to distribute \(r\) prescribed reflection pairs among the slots of
sizes \(2h_j\) and choose a reflection-invariant tree in each slot.
The empty product gives \(A(0)=1\). There are \(A(k)\) anchored
reflection-fixed frames in total. The number placing one first is
\[
 N_1=\binom{k-b/2}{b/2}\,T(b/2)\,A(k-b).
\]
To see this, the \(b\) sibling edges must form a reflection-invariant
matching disjoint from the singleton. Since \(b\) is even, the matching
omits the unique edge fixed by reflection, \(\{k,-k\}\).
Choose \(b/2\) disjoint edges on the path \(1,\ldots,k\) and take their
mirrors. After the \(b\) edges are contracted, the resulting leaves form
\(b/2\) reflection pairs in the upper tree. The unused pairs fill the remaining slots. The two first-bit
counts are therefore \(A(k)-N_1\) and \(N_1\). All integers in this
calculation have \(O(n\log n)\) bits.

At the second position \((0,2)\) every frame places zero. Swaps within
the pairs \(\{0,1\}\) and \(\{2,3\}\) preserve all sibling bits, so
the next orbit contains all four cross pairs. At least one is a nonedge,
since \(C_n\) contains no four-cycle. On \(C_7\), fix the distance pair
\(\{i,-i\}\) in the two-leaf slot. The four-leaf slot has three possible
sibling partitions. Inspecting its cross pairs gives the prefixes
\(000,000,000\) for \(i=2\) and \(100,001,001\) for each of \(i=1,3\).

\paragraph*{Completing the last rows.} Let \(R\) consist
of all required edges of a prefix on the \(n\) positions. If a vertex has degree above two or a cycle uses fewer than \(n\)
vertices, no completion is possible. A spanning cycle can be
tested directly. Otherwise \(R\) consists of \(u=n-|E(R)|\) paths,
counting isolated vertices. Each completion arranges these paths in
cyclic order and orients the nontrivial ones. Enumerate the distinct
cycles obtained this way, discard those violating prescribed zeros,
and keep only cycles \(Y\) for which \(\CF_Q(Y)=Y\).

By the orbit count under the \hyperlink{sec:base}{Sylow base bound}, a retained cycle occurs
with multiplicity \(2n/|Q\cap\Aut(Y)|\). Every element of \(Q\) fixes
the singleton slot. Of the cycle automorphisms fixing that vertex, the
only possibilities are the identity and its reflection. The multiplicity
is therefore odd precisely when the reflection preserves the binary
frame. A traversal identifies the vertices exchanged by reflection. We then
check whether the reflection preserves each level of the binary forest. This gives both child parities in
\(2^{O(u\log(u+1))}n^{O(1)}\) time by recording the next bit of each
accepted cycle. With \(r\) vertices left after the completed rows, all
\(2u\) missing incidences belong to those vertices, so \(u\le r\).
In particular the last \(O(\log n/\log\log n)\) rows admit a polynomial test.

\subsection{Initial prefixes and binary subtrees}

Let \(d\) be a power of two with \(2\le d\le B\). In each slot take
the leaf sets of the subtrees of size \(d\), or the whole slot if it is
smaller. Call these sets \emph{cells}. The first \(d-1\) positions are
\((0,1),\ldots,(0,d-1)\), all internal to a cell of the largest slot.
The set of pairs internal to these largest-slot cells is \(Q\)-invariant.
Until the initial positions have been placed, the orbit of the least
remaining position lies in this set. Each call on such an orbit reads
only its internal bits. Thus the induced graphs in the framed cells of
the largest slot determine the first \(d-1\) placed bits.

\paragraph*{A polynomial test for each fixed initial length.}
A cell type records the binary tree on \(d\) leaves and its induced graph
up to tree-preserving isomorphism. For fixed \(d\) the number of types
is independent of \(n\).
Enumerate their possible multiplicities \(m_T\) summing to \(B/d\).
To count the choices of disjoint framed cells of these types, take
\(m_T\) numbered copies of each type and count injections of their leaves
into the cycle. Within a cell require its edges and forbid its nonedges.
Between cells all pairs are unrestricted.

For type \(T\), add a subset of the forbidden edges to its required edges
and give the term the inclusion--exclusion sign of that subset. A vertex
of degree above two precludes an embedding. So does a cycle, since
\(B<n\). In every other term the graph is a union of paths. Record
\(x^c y^a\) with the corresponding sign, where \(c\) counts all path
components, including isolated vertices, and \(a\) counts the nontrivial
ones. Let \(P_T\) be the resulting polynomial. In \(\prod_T P_T^{m_T}\), multiply the coefficient of \(x^c y^a\) by
\[
 n(c-1)!\,2^a\binom{n-B+c-1}{c-1}.
\]
Choose a clockwise direction. A designated path starts at one of \(n\) cycle positions.
Order the others around the cycle, orient the nontrivial paths and
distribute the \(n-B\) unused vertices among the gaps. Summing gives the
injection count. Divide this integer by \(\prod_T m_T!\,|\Aut(T)|^{m_T}\). The factorials
forget the ordering of cells of the same type. The automorphism factors
forget the isomorphisms to their numbered representatives. The product has degree at most \(B\) in each variable, and its coefficients
have polynomial bit length.

It remains to sum over the trees joining the cells. Permuting equal-type
cells preserves the test by locality. In each type, partition its \(m_T\)
cells into batches of the sizes in the binary expansion of \(m_T\).
Put a fixed binary tree on each batch. The product of the automorphism
groups of these trees is a \(2\)-group, so only its fixed upper trees
contribute to parity. Each batch must occur intact with its prescribed
hierarchy. A transposition fixing all other leaves forces the two leaves
it exchanges to be siblings. Inductively a swap of two already identified
subtrees fixing everything outside them forces them to be siblings as
well. Conversely, every balanced tree joining the intact batches is fixed.

If there are \(s\) types, at most \(s\) new batches enter at each size
and at most \(s\) trees arrive from the previous size. Pair these trees
in all possible ways at each level. At most \((2s)!\) choices per level
over \(O(\log n)\) levels give polynomially many trees for fixed \(d\).
For each upper tree, set the edges outside the cells to zero and run
placement. Multiply the parity of the number passing the prefix test by
the number of choices of cells. The forest outside
the largest slot has an odd number of choices, the index of its binary
tree group in the symmetric group. Summing over the multiplicities gives the required parity. The degree of
the running-time polynomial depends on \(d\).

\paragraph*{A cancellation for growing cells.}
Suppose \(B>2d\). Consider anchored frames fixed by reflection in which
reflection exchanges the two children of the largest tree. For every
prefix of length at most \(d-1\), an even number of these frames give
that prefix. This holds also when \(d\) grows with \(n\).

To prove this, group the distance labels \(1,\ldots,k\) according to the
reflection orbit of the cell containing \(i\). The cell containing \(-i\)
lies in the same orbit. Fix the resulting partition of distance labels.
For each maximal consecutive interval in one part, interchange \(i\) and
\(-i\) on that interval. These permutations preserve the partition and
commute with reflection. They preserve the anchored family under study,
including the condition on the two children of the largest tree.

A flip can change adjacencies only across an interval boundary. There the
two consecutive labels are in different cell orbits, so none of the
changed pairs is internal to a cell. Each flip preserves the set of edges incident with \(0\) and fixes the
edge \(\{k,-k\}\) setwise. By locality, each prefix
test is constant on orbits of the \(2\)-group generated by the flips.
It suffices to consider frames fixed by every flip. If reflection
exchanges two cells in such a frame, their distance labels form a single interval. Otherwise a flip of one
maximal interval would move only part of a cell to its mirror.
Flipping the whole interval exchanges these two cells and fixes all
other leaves. The cells must be siblings, since a swap at their least common ancestor
moves every leaf below it. Hence reflection preserves
every subtree of size at least \(2d\) setwise. When \(B>2d\) it cannot
exchange the two root children, which proves the cancellation.
For \(B\ge8\) we may take \(d=B/4\).

On \(C_{17}\) the cancellation fails at a later reached prefix. Removing the
singleton leaves a path with a unique perfect matching. Contracting its
edges to eight leaves reduces the check to the \(315=8!/2^7\) balanced
trees on those leaves. Keep the \(51\) preserved by path reversal and
apply the \hyperlink{sec:upper}{canonical placement algorithm} to each expanded tree.
At the reached prefix \(1\,0^{10}\), the zero and one counts are
\((21,1)\) for exchanged halves and \((27,0)\) for preserved halves.
Together these give \((48,1)\), so the exchanged halves determine the
next bit.

\subsection{Width of a representation on adjacency bits}

Let \(S_s\) be the set of distinct placed cycles extending \(s\) with
odd frame multiplicity. A \emph{pointwise factorisation} represents
its indicator as \(\sum_z\prod_i f_i(Y,z)\) over \(\mathbb F_2\) for
every assignment of the adjacency bits \(Y_e\), with auxiliary Boolean
variables \(z\). Summing out all but the next adjacency bit then gives
the two child parities.

In a tree decomposition each node carries a set of variables, or
\emph{bag}. Every factor has its variables together in a bag, and
the bags containing each variable form a connected subtree. An
elimination order of width \(w\) yields bags of size at most \(w+1\).

\begin{namedproposition}[Width at the prefix \(00\)]
For \(n=2^a-1\ge7\), the prefix \(00\) is reached on \(C_n\). Every
pointwise factorisation of \(S_{00}\) on its adjacency bits has bag
size \(\Omega(\sqrt n)\).
\end{namedproposition}
\begin{proof}
\emph{Size of the surviving family.} Here \(k=2b-1\). In the first-bit count, \(b/2\) is a power of two.
Writing \(h=b/2\), the binomial coefficient is \(\binom{3h-1}{h}\).
It is even because, over \(\mathbb F_2\), the coefficient of \(x^h\) in
\((1+x)^{3h-1}=(1+x^{2h})(1+x)^{h-1}\) vanishes.
The root count is odd, hence the algorithm chooses zero. The second bit
always vanishes by the sibling-swap argument above.

There are at least \(k!/2^k\) distinct members of \(S_{00}\).
Make every reflection pair \(\{i,-i\}\) a sibling pair and choose an
arbitrary binary frame on the \(k\) pairs. The number of choices is
\(k!/2^{v_2(k!)}\). Only one reflection pair is a cycle edge, so the
largest slot contains a sibling nonedge and the placed prefix is \(00\).
Two anchored frames have the same placed string precisely when a cycle
automorphism fixing \(0\) transports one to the other. Both such
automorphisms fix every constructed frame. The strings are therefore
distinct, each represented once among anchored fixed frames.

\emph{A counting bound for fixed degrees.} Let
\(\mathcal F\) be any family on \(m\) possible edges with a common
degree vector \(D\) and \(q\ge1\) edges per graph. If its indicator has
bag size \(W\), put \(b'=W+\lceil\log_2(2m+1)\rceil\). We claim that
\(\lvert\mathcal F\rvert\le (8b')^q\).

Attach a leaf for each visible bit to a bag containing it, using an
equality to read that bit. Split high-degree bags into binary trees of
identical bags, prune branches with no visible leaves, and suppress unary
nodes. In the resulting binary tree the \(m\) leaves are the possible
edge positions. Across any cut, the matrix of the indicator with rows
and columns indexed by assignments on opposite sides has rank at most
\(2^W\). Fixing the shared bag variables separates the factors. Summing
over their assignments gives at most \(2^W\) outer products, also over
\(\mathbb F_2\). A leaf cut has rank at most two.

Each degree vector \(u\) realised on the descendant side requires vector
\(D-u\) on the other, with degrees taken over the integers. Group rows
and columns by these vectors. Distinct realised vectors give disjoint
nonzero diagonal blocks, so there are at most \(2^W\) of them per cut.

To encode a graph from the family, mark the \(q\) leaves corresponding
to its edges. Work upwards, detaching
a nonroot subtree as soon as it contains at least \(b'\) still-attached
marks. Each child passed up fewer than \(b'\) marks, so every resulting
piece has at most \(2b'\) marks. At most \(\lfloor q/b'\rfloor\) cuts
were made. Record each cut and its degree vector for the entire original
descendant subtree. Padding the list to \(\lfloor q/b'\rfloor\) entries with a blank symbol gives at most
\(((2m+1)2^W)^{\lfloor q/b'\rfloor}\le2^q\) possible values.

Subtracting the vectors at the cuts immediately below a piece from its
subtree vector (using \(D\) at the root) gives the degrees inside the
piece. If it has \(j>0\) marked edges, those degrees specify \(2j\)
labelled edge ends. At most \((2j)^j\le(4b')^j\) pairings need be
considered. A piece with no edges has one possible graph. Multiplication
over the pieces gives at most \((4b')^q\) graphs per record, proving
the claim. Now take \(m=\binom n2\) and \(q=n\). The two bounds give
\(8b'\ge(k!/2^k)^{1/n}=\Omega(\sqrt n)\). Subtracting the logarithmic
term gives the stated lower bound on \(W\).
\end{proof}

The fixed-length algorithm computes both child parities at \(00\) in
polynomial time. It does so without representing membership in \(S_{00}\)
on all adjacency bits.

Whether both child parities can be computed in polynomial time at every
reached prefix remains open even for odd cycles. A polynomial-time test for
the zero child at every reached prefix on arbitrary graphs would give a
logic capturing PTIME by evaluation on the canonical copy.

\end{document}